\documentstyle[aps,epsf]{revtex}
\def \be   {\begin{equation}}
\def \ee   {\end{equation}}
\def \l {\label}
\begin{document}
\input epsf
\baselineskip=25pt
\title{Discrete interactions and the Pioneer anomalous acceleration}
\author{Manoelito M de Souza}
\address{{Departamento de
F\'{\i}sica - Universidade Federal do Esp\'{\i}rito Santo\\29065.900 -Vit\'oria-ES-Brazil}\thanks{ E-mail: manoelit@verao.cce.ufes.br}}
\date{\today}
\maketitle
\begin{abstract} The dominant contributions from a discrete gravitational interaction produce the standard potential as an effective continuous field. The sub-dominant contributions are, in a first approximation, linear on $n$, the accumulated  number of (discrete) interaction events along the test-body trajectory. For a nearly radial trajectory $n$ is proportional to the traversed distance and its effects may have been observed as the Pioneer anomalous constant radial acceleration, which cannot be observed on the nearly circular planetary orbits.
\end{abstract}
\begin{center}
\end{center}
The constant anomalous radial acceleration observed \cite{gr-qc/0104064,Pioneer} in the Pioneer 10 and 11 spacecrafts defies explanation. A Yukawa-type interaction, in addition to Newtonian gravity, would not be compatible with the planetary ephemeris. In this letter we show how a discrete gravity naturally fits both the Pioneer and the planetary data and predicts a distance dependence of this anomalous acceleration to be observed with an accuracy improvement on its determination. .
The discrete field concept \cite{hep-th/9610028,gr-qc/981040} relies on the assumption of an absolutely discrete world - made of discretely interacting pointlike objects, the discrete fields. Any continuity, except of the spacetime, is just apparent, a matter of scale. Field theory becomes, in a way, simpler and finite. The standard continuous-field theories, including Classical Electrodynamics \cite{hep-th/9610028} and General Relativity \cite{gr-qc/981040,hep-th/0103218}, are retrieved in terms of effective averaged fields. Then discrete field theories encompass the success of the continuous-field theories and explain their limits. The worldline of a discretely interacting physical object is continuous but  (piecewise) differentiable only between consecutive interaction events. The concept of acceleration makes sense then only as an approximative limiting continuous description in a range where this approximation is allowed and defined by the impossibility of an experimental resolution. We must deal, instead, with sudden discrete changes of velocity at the interaction events. Interaction strengths are then parameterized by the time intervals $\Delta t_{i}$ between consecutive interaction events and by the sudden change of velocity $\Delta v_{i}$ there (spin has not been considered yet). The ratio $\frac{\Delta v_{i}}{\Delta t_{i}}$ has no special   meaning. Outside this range, in the very small and in the very large distance limits, where the ratio $\frac{\Delta v_{i}}{\Delta t_{i}}$ can no longer fit the acceleration definition, discrepancies can be observed between a continuous and a discrete descriptions. See the Figure 3 of \cite{hep-th/0103218}. Here we will consider the very-large-distance limit where the fields are weak and the velocities small so that relativistic corrections can be neglected. We will be concerned with the apparent anomalous constant acceleration detected in the Pioneer X and XI, which according to the authors of references \cite{gr-qc/0104064,Pioneer}, has found no consistent theoretical or systematic explanation so far. ``The anomalous acceleration is too large to have gone undetected in planetary orbits, particularly for Earth and Mars"\cite{gr-qc/0104064}. Here we show that it can be the evidence of second-order contributions from a discrete gravitational interaction. They are proportional to $n$, an integer orderly labelling the interaction events along the probe worldline and it is, in a first approximation, proportional to $\Delta r=r_{n}-r_{0}$ for a radial worldline but, of course, not for a circular one.

The much simpler case of a radial motion with a non-relativistic axially symmetric interaction (where the effective acceleration is inversely proportional to the distance) has been considered in \cite{hep-th/0103218}. Here we will consider a non-relativistic radial motion with a discrete interaction whose effective description is an acceleration field that decreases with the inverse of the squared distance. The following two possible alternative descriptions will be discussed:
\begin{itemize}
\item
Alternative I is based on an assumed universal quantum of interaction causing discrete changes of velocity 
\be
\l{dv1}
\Delta v_{i}\equiv\Delta_{I},
\ee
where $\Delta_{I}$ represents a universal constant interaction. Rigourously, instead of a change of velocity , we should have a change of four momentum, representing the energy-momentum carried by the quantum of (discrete) interaction, but weak fields and low velocities justify this non-relativistic simplification. The rest frame of a central source, a mass $M$ in the case of gravity, is then assumed. The time interval $\Delta t_{i}$ between two consecutive interactions must then be given by
\be
\l{dti1}
\Delta t_{i}=\alpha_{I} r^{2}_{i},
\ee
where $\alpha_{I}$ is another constant,
in order to reproduce the observed effective Newtonian acceleration.
\item Alternative II assumes
\be
\l{dti2}
\Delta t_{i}=\alpha_{II} r_{i},
\ee
which has a nice intuitive interpretation for $\Delta t_{i}$  as the two-way flying-time of the interaction quantum; it requires
\be
\l{dv2}
\Delta v_{i}\equiv\frac{\Delta_{II}}{r_{i}}
\ee for the dynamics.
See the reference \cite{hep-th/9610145} for a physical interpretation.
\end{itemize} This second alternative clearly corresponds to an effective description which is denounced by the singularity (infinity) at the origin. It does not fit the discrete-field philosophy which implies finite interactions ever. It is included here, together with the first one, for the sake of completeness because both alternatives reproduce the Newtonian gravity as an effective field and the Pioneer acceleration as a second order effect not included in the standard continuous descriptions. Only the first alternative is physically acceptable as a description of elementary processes  since the validity of the second one is restricted to large distances.
For initial conditions taken as $$r(t_{0})= r_{0};\qquad v(t_{o})=v_{0},$$ the next interaction will occur at 
\be
\l{Dt0}
t_{1}= t_{0}+\Delta t_{0},
\ee
with 
\be
v(t_{1})\equiv v_{1}=v_{0}-\Delta v_{1},
\ee
\be
\l{r1}
r(t_{1})\equiv r_{1}= r_{0}+v_{0}\Delta t_{0},
\ee
as we are neglecting relativistic corrections, and there is free propagation between  consecutive interaction events.  At the $n^{th}$ interaction,
\be
\l{vn}
v_{n}=v_{0}-\sum_{i=1}^{n-1}\Delta v_{i},
\ee
\be
\l{rn}
r_{n}=r_{0}+\sum_{i=1}^{n-1}\Delta v_{i}\Delta t_{i}.
\ee
The Eqs. (\ref{dv1},\ref{dti1}) or (\ref{dti2},\ref{dv2}) replace the differential equations of the continuous fields and the $(n+1)$-term finite series of Eqs. (\ref{vn},\ref{rn}) replace their respective continuous solutions.

The alternative I leads to
\be
\l{vn1}
v_{n}=v_{0}-n\Delta _{I},
\ee
\be
\l{rn1}
r_{n}=r_{n-1}+\alpha_{I}r^{2}_{n-1}v_{n-1},
\ee
which, recursively, produces the following finite series
\be
\l{rnrj}
\frac{r_{n}}{r_{j}}=1+\alpha_{I}r_{j}\sum_{i=1}^{n-j}v_{n-i}+2!\alpha_{I}^{2}\sum_{i_{1}=2}^{n-j}\sum_{i_{2}=1}^{i_{1}-1}v_{{n}-i_{1}}v_{n-i_{2}}+\alpha_{I}^{3}r_{j}^{3}{\Big{\{}}\sum_{i_{1}=2}^{n-j}\sum_{i_{2}=1}^{i_{1}-1}v^{2}_{n-i_{1}}v_{n-i_{2}}+3!\sum_{i_{1}=3}^{n-j}\sum_{i_{2}=2}^{i_{1}-1}\sum_{i_{3}=1}^{i_{2}-1}v_{n-i_{1}}v_{n-i_{2}}v_{n-i_{3}}{\Big{\}}}+\dots
\ee
These successive sums may become quite involved as n usually is a huge number and so the adoption of a systematic approach is very convenient. 
We use 
\be
 {n\choose 0}=\cases{0,&if $n<k$;\cr
\cr
_1,&if $k=0$;\cr
\cr
\frac{n!}{(n-k)!k!},&if $n>k>0$.\cr}
\ee
 with
\be
\l{eeq}
\sum_{i=0}^{n-1}{i\choose k}={n\choose k+1},
\ee
$$\sum_{i=1}^{n}\equiv\sum_{i=0}^{n-1}+\delta_{i}^{n}-\delta_{i}^{0},$$
$$\sum_{i=0}^{n-1}i{i\choose k}=(k+1){n\choose k+2}+k{n\choose k+1},$$
$$\sum_{i=0}^{n-1}i^K{i\choose k'}=\frac{k+k'}{k'!}{n\choose k+k'+1}+{\hbox{smaller order terms}},$$
and other results derived from these \cite{gr-qc/0106047} for a systematic writing of each term of this series in terms of combinatorials ${n\choose k}$. 

Keeping only the dominant contribution from the combinatorials of each term is a very good approximation when $n$ is a very large integer.  In a such approximation the exact Eq. (\ref{eeq}) is replaced by
$$\sum_{i}^{n-p}{i\choose k}\approx{n\choose k+1}\approx\frac{n^{k+1}}{(k+1)!},$$
valid for each $\sum$ in Eq. (\ref{rnrj}), $p<n$. Then the $k^{th}$ term of this equation becomes
\be
k!\alpha_{I}^{k}r_{j}^{k}\sum_{i_{1}=k}^{n-j}\sum_{i_{2}=k-1}^{i_{1}-1}\dots\sum_{i_{k}=1}^{i_{k-1}-1}v_{n-i_{1}}v_{n-i_{2}}\dots v_{n-i_{k}}
\approx[\alpha_{I}r_{j}n(v_{n}+\frac{n\Delta_{I}}{2}]^{k}
=[\frac{\alpha_{I}r_{j}}{\Delta_{I}}(\frac{v^{2}_{0}-v^{2}_{n})}{2}]^{k},
\ee
where use was made of Eq. (\ref{vn1}).
 Then we have from Eq. (\ref{rnrj})
\be
\frac{r_{n}}{r_{0}}=\sum_{k=0}^{n}\quad[\frac{\alpha_{I}r_{j}}{\Delta_{I}}(\frac{v^{2}_{0}-v^{2}_{n}}{2})]^{k},
\ee
which can be approximated, as $n$ is a large number, by
\be
\l{xn}
\frac{r_{n}}{r_{0}}\approx\frac{1}{1-\frac{\alpha_{I}r_{0}}{\Delta_{I}}\frac{(v_{0}^2-v_{n}^2)}{2}}.
\ee
 This result is equivalent to
\be
\l{energy}
\frac{v_{n}^2}{2}-\frac{\Delta_{I}}{\alpha_{I}}\frac{1}{r_{n}}=const,
\ee
which retrieves the Newtonian potential $U(r)=-\frac{GM}{r}$, as an effective field and implies, as expected from Eqs. (\ref{dv1},\ref{dti1}), that 
\be
\frac{\Delta_{I}}{\alpha_{I}}=GM.
\ee
Energy is always conserved but $U(r)$ is valid as the expression of a potential energy only as an asymptotic idealized limit of $n\rightarrow\infty$. The discrimination between  the continuous and the discrete interaction-descriptions must experimentally be decided by the detection or not of effects from the neglected smaller contributions. The inclusion of the first-order (in $\alpha_{I}$) non-dominant contribution from Eq. (\ref{rnrj}) changes the Eq. (\ref{xn}) to
\be
\l{xns}
\frac{r_{n}}{r_{0}}\approx\frac{1}{1-\alpha_{I}r_{0}n(v_{n}+n\Delta_{I}/2)}+\alpha_{I}r_{0}n\frac{\Delta_{I}}{2}+{\cal {O}}(\alpha_{I}^2),
\ee
which leads to
\be
(\frac{v_{n}^2}{2}-\frac{\Delta_{I}}{\alpha_{I}}\frac{1}{r_{n}})-(\frac{v_{0}^2}{2}-\frac{\Delta_{I}}{\alpha_{I}}\frac{1}{r_{0}})=n\frac{\Delta_{I}^2}{2}+{\cal{O}}(\alpha_{I}).
\ee
It is significative that the extra interaction, up to first order terms in $\alpha_{I}$, is directly proportional to $n$. As we are considering a probe on a radial trajectory, $n$ , in this order of approximation, is roughly proportional to $\Delta r\equiv r_{n}-r_{0}$, the traversed distance. This leads to the observed Pioneer acceleration. Furthermore, for nearly circular trajectories, if a similar term appears, it should not be proportional to $\Delta r$. So, there is no conflict with the planetary data where this acceleration is not observed. 
 
From Eq. (\ref{rnrj}),
\be
\frac{r_{n}}{r_{0}}=1+\alpha_{I}r_{0}n(v_{n}+\frac{n\Delta_{I}}{2})+{\cal{O}}(\alpha_{I}^2).
\ee
Then, using Eq. (\ref{vn1}),
\be
\frac{dn}{dr_{n}}=\frac{1}{\alpha_{I}r_{0}^{2}v_{n}},
\ee

\be
a_{P}=\frac{\Delta_{I}^2}{2\alpha_{I}r_{0}^{2}v_{n}}=\frac{a_{1}\Delta_{I}}{2v_{n}},
\ee
and
\be
\frac{da_{P}}{dr_{n}}=\frac{a^{2}_{1}\Delta_{I}}{2v^{3}_{n}}=\frac{a_{1}a_{P}}{v^{2}_{n}},
\ee
where $a_{P}$ is the extra acceleration and $a_{1}$ is the value of the standard acceleration at $r_{0}=1UA$. 
If we take \cite {gr-qc/0104064} $a_{P}\simeq(8,74\pm0.94)\times10^{-10}m/s^{2}$ we have
$$\Delta_{I}=2v_{n}\frac{a_{P}}{a_{1}}\approx v_{n}\times10^{-8,}$$
$$\alpha_{I}\approx{v}_{n}10^{-28}\quad[MKS]$$ and $$\frac{da_{P}}{dr_{n}}\approx\frac{10^{-11}}{v^{2}_{n}}\quad[MKS]$$
For (astronomically) large distances $n$, and therefore $v_{n}$, changes very slowly so that taking $a_{P}$ as a constant is a good approximation in face of the present accuracy on the $a_{P}$ experimental determination. But if this can be improved it must be found, according to these results, that $a_{P}$, in this approximation, increases with the distance as $v_{n}$ decreases.

The alternative II leads to
\be
\l{rnII}
r_{n}=r_{n-1}+v_{n-1}\Delta t_{n-1}=(1+\alpha_{II} v_{n-1})r_{n-1}=r_{0}\amalg_{i=0}^{n-1}(1+\alpha_{II} v_{i}),
\ee
 with 
\be
\l{vj}
v_{j}=v_{0}-\sum_{i=1}^{j}\frac{\Delta_{II}}{r_{i}}.
\ee
Then
\be
\l{rn0}
\frac{r_{n}}{r_{0}}= 1+\alpha_{II}\sum_{i_{1}=0}^{n-1}v_{i_{1}}{\Big\{}1+\alpha_{II}\sum_{i_{2}=i_{1}+1}^{n-1}v_{i_{2}}{\Big\{}+\dots+\sum_{i_{n-1}=i_{n-2}+1}^{n-1}v_{n-1}{\Big\}}\dots{\Big\}},
\ee
\be
\l{vnj}
v_{j}=v_{0}-\frac{K}{r_{n}}\sum_{i=1}^{j}\frac{r_{n}}{r_{i}}=v_{0}-\gamma_{n}{\Big\{}1+\alpha_{II}\sum_{i_{1}=j}^{n-1}v_{i_{1}}{\Big\{}1+\alpha_{II}\sum_{i_{2}=i_{1}+1}^{n-1}v_{i_{2}}{\Big\{}+\dots+\sum_{i_{n-1}=i_{n-2}+1}^{n-1}v_{n-1}{\Big\}}\dots{\Big\}}{\Big\}},
\ee
with $\gamma_{n}=\frac{K}{r_{n}}$. Calculation details are being spelled at \cite{gr-qc/0106047}.

Keeping again only the dominant contribution from each term in each series 
\be
\l{svn}
v_{n}={\Big \{}v_{0}-n\gamma_{n}{\Big \}}-\alpha_{II}{\Big \{}n^2\gamma_{n}(\frac{v_{0}}{2}-\frac{n\gamma_{n}}{3}){\Big \}}-\alpha_{II}^2{\Big \{}\frac{n^3\gamma_{n}}{3}(\frac{v_{0}^2}{2}-\frac{5}{4}v_{0}n\gamma_{n}+\frac{4}{5}n^2\gamma_{n}){\Big \}}-\dots
\ee
and
\be
\l{sxn}
\frac{r_{n}}{r_{0}}=1+\alpha_{II}{\Big \{}n(v_{0}-\frac{n\gamma_{n}}{2}){\Big \}}+\alpha_{II}^2{\Big \{}\frac{n}{2}(v_{0}^2-\frac{6}{3}nv_{0}\gamma_{n})+\frac{2n^2\gamma_{n}^2}{3}){\Big \}}+\dots
\ee
Introducing the following notation
\be
\l{vs}
v_{n}=\sum_{s=0}^{s=n}\alpha_{II}^s V^{(s)}_{n},
\ee 
for the terms of the series (\ref{svn}) we can use it for writing the series (\ref{sxn}) as
\be
\l{xsn1}
\frac{r_{n}}{r_{0}}=1+\alpha_{II}{\Big \{}\frac{v_{0}^2-V_{n}^{(0)^{2}}}{2\gamma_{n}}{\Big \}} -\alpha_{II}^2{\Big \{}\frac{2v_{0}V_{n}^{(1)}}{2\gamma_{n}}{\Big \}}-\alpha_{II}^3{\Big\{} 
\frac{V_{n}^{(1)^{2}}+2V_{n}^{(0)}V_{n}^{(2)}}{2\gamma_{n}}{\Big \}}-\dots=1+\frac{\alpha_{II}}{\gamma_{n}}(\frac{v_{0}^2-v_{n}^{2}}{2}),
\ee
which  is equivalent to Eq. (\ref{energy}),  an expression of energy conservation in terms of the effective Newtonian potential. 
Therefore, the inclusion of just the first non-dominant contribution, leads the Eq. (\ref{sxn}) to 
\be
\l{KUn}
\frac{r_{n}}{r_{0}}-1-\frac{\alpha_{II}}{\gamma_{n}}(\frac{v_{0}^2-v_{n}^{2}}{2})=-\frac{\alpha_{II}n\gamma_{n}}{2}+{\cal{O}}(\alpha^{2}_{II}),
\ee
which can be rearranged as
\be
\l{Xlinha}
(\frac{v_{0}^2}{2}-\frac{K}{\alpha_{II}}\frac{1}{r_{0}})-(\frac{v_{n}^2}{2}-\frac{K}{\alpha_{II}}\frac{1}{r_{n}})=-\frac{n\gamma^{2}_{n}}{2}+{\cal{O}}(\alpha_{II})\approx-\frac{n\Delta_{II}^{2}}{2r^{2}_{0}}+{\cal{O}}(\alpha_{I}).
\ee
This reproduces, at this  order of approximation, the same results of alternative I but with 
\be
\frac{\Delta_{II}}{r_{0}}\approx\Delta_{I}\approx2v_{n}\times10^{-8}\quad[MKS]
\ee
With $r_{0}\sim10^{11}m$, this implies a $\Delta_{II}\sim v_{n}\times10^{3}m^{2}/s$ which excludes the second alternative from a viable description of elementary processes  as Eq. (\ref{dv2}) then is valid only for large distances.


\begin{thebibliography}{10}
\bibitem{gr-qc/0104064}J.D. Anderson, P. A. Laing, E. L. Lau, A. S. Liu, M. M. Nieto, S. G. Turyshev, {\it{Study of the anomalous acceleration of Pioneer 10 and 11}}, gr-qc/0104064. 
\bibitem{Pioneer}J.D. Anderson, P. A. Laing, E. L. Lau, A. S. Liu, M. M. Nieto, S. G. Turyshev, Phys. Rev. Lett. 81, 2858(1998). gr-qc/9808081; {\it{The apparent anomalous, Weak, Long-Range Acceleration of Pioneer 10 and 11}}, S.G. Turyshev, J. D. Anderson, P. A. Laing, E. L. Lau, A. S. Liu, M. M. Nieto. gr-qc/9903024.
\bibitem {hep-th/9610028} M. M. de Souza, J. of Phys A: Math. Gen. 30 (1997)6565-6585; {\it {Classical Electrodynamics and the Quantum Nature of Light}}, hep-th/9610028; {\it Conformally symmetric massive discrete fields}, hep-th/0006237. 
\bibitem{gr-qc/981040} M. M. de Souza, Robson N. Silveira,  Class. \& and Quantum Gravity, vol 16, 619(1999),({\it{Discrete field and General Relativity}}, gr-qc/981040).
\bibitem{hep-th/0103218} M. M. de Souza, {\it{Discrete fields, general relativity, other possible implications and experimental evidences}}, hep-th/0103218.
\bibitem{hep-th/9610145} M. M. de Souza, {\it Classical fields and the quantum concept}. hep-th/9610145.
\bibitem{gr-qc/0106047}M. M. de Souza,{\it Discrete interactions and the Pioneer anomalous acceleration: Alternative II.}, gr-qc/0106047.
\end{thebibliography}
\end{document}